\documentclass{article}
\begin{document}
\title{Comment on `Non-realism: deep thought or a soft option?', by N. Gisin}
\author{Michael J. W. Hall\\
Theoretical Physics, RSPE, Australian National
University, \\Canberra ACT 0200, Australia}
\date{}
\maketitle

% USE THE FOLLOWING INSTEAD OF ABOVE FOR REVTEX (SEE ALSO PACS BELOW)
%\documentclass[print, showpacs, aps, draft]{revtex4}
%\usepackage{bm}

%\begin{document}
%\title{Prior information: how to beat the standard joint-measurement
%uncertainty relation}
%\author{Michael J. W. Hall}
%\affiliation{Theoretical Physics, IAS, \\ Australian National
%University,\\
%Canberra ACT 0200, Australia}
%\date{}

\begin{abstract}
It is briefly demonstrated that Gisin's so-called `locality' assumption [in http://arxiv.org/abs/0901.4255] is in fact equivalent to the existence of a local deterministic model.  Thus, despite Gisin's suggestions to the contrary, `local realism' in the sense of Bell is built into his argument from the very beginning.  His `locality' assumption may more appropriately be labelled `separability'.  It is further noted that the increasingly popular term `quantum nonlocality' is not only misleading, but tends to obscure the important distinction between no-signalling and separability.  In particular, `local non-realism' remains firmly in place as a {\it hard} option for interpreting Bell inequality violations. Other options are briefly speculated on.
\end{abstract}

%USE FOR REVTEX
%\pacs{03.65.Ta}
%\maketitle

\section{Introduction}

Gisin has recently concluded that ``{\it all violations of Bell's inequality should be interpreted as a demonstration of nonlocality}'' \cite{gisin}. However, his definition of `locality' is in fact equivalent to the existence of a local realistic model (in the original sense of Bell \cite{bell}), as is shown in section 2 below.   Hence, his refutation of a role for realism -  ``{\it an alleged alternative between nonrealism and nonlocality is annoying}''  \cite{gisin} - is not logically sustainable.  

It is therefore suggested in section 3 that Gisin's `locality' assumption would be better labelled as  `separability'.  For similar reasons, it is further suggested that use of the misleading term `quantum nonlocality', which is becoming popular in the literature, be avoided, as it implicitly suggests there is, {\it as a matter of fact}, some mysterious faster-than-light influence operating in quantum systems.  While the possibility of such an influence cannot be denied, there is no physical basis at present for its assertion as a consequence of measurable Bell-inequality violations.  Indeed, as noted in section 4, the principle of Occam's razor suggests the opposite, since the standard Hilbert space model of quantum correlations is local but not realistic. 

Some definitions are suggested in section 4 to clarify debates concerning locality and realism in quantum mechanics, and some brief speculations are made in closing as to the nature of the distinction between no-signalling and separability.

\section{Equivalence of Gisin's `locality' with local realism}

Gisin identifies `locality' as the assumption that the correlations measured by two observers can be modelled by joint probabilities of the form \cite{gisin}
\begin{equation} \label{gloc}
p(a,b|x,y) = \int d\lambda\,\rho(\lambda) \,p(a|x,\lambda) \,p(b|y,\lambda)  ,
\end{equation}
where $a$ and $b$ label respective measurement results for detector arrangements $x$ and $y$, and $\lambda$ is some variable describing the physical state (eg, a quantum state, possibly supplemented by `hidden variables'), with corresponding probability density $\rho(\lambda)$.

Now, a {\it local deterministic} model may be defined as one for which the quantities $p(a|x,\lambda)$ and $p(b|y,\lambda)$ in equation (1) are restricted to take the values $0$ and $1$.  Such models are also commonly called {\it local realistic models}, following Bell's original terminology \cite{bell}.  However, while it might initially appear that models satisfying equation (1) form a broader class of models than the class of local realistic models, the difference is purely cosmetic.  In particular, one has the following result:

{\bf Theorem:} {\it For any set of measurement pairs $\{(x,y)\}$, there exists a model satisfying equation (1) if and only if there exists a local realistic model}.

{\it Proof:}  Suppose first one has a model satisfying equation (1).  Choose some ordering of the possible results, $\{a_j\}$, for each measurement (eg, for Stern-Gerlach measurements on spin-1/2 particles, one could take the natural ordering \{ spin aligned with magnetic field, spin anti-aligned with magnetic field\}). Now define a corresponding local deterministic model via (i) the variable 
\[ \tilde{\lambda}\equiv (\lambda,\alpha,\beta) , \] 
where $\alpha$ and $\beta$ are uniformly and independently distributed over the interval $[0,1)$; (ii) the corresponding probability density
\[ \tilde{\rho}(\tilde{\lambda}) = \tilde{\rho}(\lambda,\alpha,\beta) := \rho(\lambda),  \]
for $\tilde{\lambda}$; and (iii)  deterministic probabilities
\[ 
p(a_j|x,\tilde{\lambda}) := 1~~{\rm for~~} \alpha \in \left[\sum_{i< j} p(a_i|x,\lambda)\right.,\left. \sum_{i\leq j} p(a_i|x,\lambda)\right),\]
\[
p(b_k|y,\tilde{\lambda}) := 1~~{\rm for~~} \beta \in \left[\sum_{i< k} p(b_i|y,\lambda)\right.,\left. \sum_{i\leq k} p(b_i|y,\lambda)\right) , 
\]
and equal to zero otherwise.  It is trivial to check that, by construction, for any pair of measurements $x$ and $y$ one then has
\[ p(a_j,b_k|x,y) = \int d\tilde{\lambda} \,\tilde{\rho}(\tilde{\lambda})\, p(a_j|x,\tilde{\lambda})\, p(b_k|y,\tilde{\lambda}) . \]
Hence, there is a local realistic model as claimed.  The converse is trivial (any local realistic model satisfies equation (1) by definition), and the theorem is proved. QED. $\diamond$

Since either model reproduces precisely the same observable joint probabilities, $p(a_j,b_k|x,y)$, it follows that any measurements of these probabilities {\it per se}, eg, in Bell-inequality tests, cannot distinguish between the models.  Hence, one cannot claim that all local realistic models are experimentally refuted without also claiming that all models satisfying equation (1) are refuted, and vice versa.   Equation (1) is thus both formally {\it and} empirically equivalent to `local realism'.  

It might be argued, of course, that nature could in `reality' be such that equation (1) is satisfied, but with no variables such as $\alpha$ and $\beta$ existing to provide a deterministic model.  However, while such an argument would allow `nonrealism' to be maintained in the interpretation of equation (1), it would do so via mere metaphysical assertion, with no testable physical consequences - similar in spirit to discussing how many angels can dance on the head of a pin.  

The above theorem is a simple generalisation of existing results in the literature for single measurements \cite{bell2,hall}. Note that the assumed ordering means that the model is (locally) contextual \cite{bell2,hall}.  Fine has previously used a rather different (nonlocally contextual) construction to obtain a form of the theorem for the case of four measurement pairs \cite{fine}, which can be generalised to the case of a countable set of measurement pairs \cite{fine2}.  In contrast, the above theorem applies to {\it arbitrary} sets of measurement pairs, such as spin measurements in all possible directions.

It is concluded that Gisin's puzzlement over ``{\it possible hiding places for realism}''  \cite{gisin}, is easily resolved - it is hidden, right from the start, in his so-called `locality' assumption (1) !  Clearly, this assumption has more to it than meets the eye, and a more suitable label is required.  This is considered below.

\section{Let's call a spade a spade}

One sentiment of Gisin's that I do strongly agree with is expressed in the first paragraph of \cite{gisin}: ``{\it why should one use the word[s] local realism rather than local determinism?}''  The second term is, after all, far less loaded with metaphysical luggage.  The popularity of the first term, despite the vagueness of `realism', is due of course to its introduction by Bell in his famous paper \cite{bell} (no doubt as a homage to the discussion of `elements of reality' in the similarly famous paper of Einstein, Podolsky and Rosen \cite{epr}).  

However, as follows from the previous section, one must similarly ask ``{\it why should equation (1) be labelled as `locality', when it is certainly much more than this?}''.  In fact, historically, equation (1) has had many previous labels, a number of them being more apposite.

For example, Gisin also refers to equation (1) as `conditional independence', and it has been previously been called `objective locality' \cite{ch}, `local explicability' \cite{bell3}, the combination of `parameter independence' (no-signalling) with `outcome independence' \cite{jarrett}, and the combination of  no-signalling with `classical logic' \cite{hall}.  

In contrast, I would like promote an alternative term, {\it `separability},' for the property corresponding to equation (1).  Its usage dates back, at least, to discussions by d'Espagnat \cite{despagnat}, and implicitly incorporates the notion that joint correlations of this form can be decomposed or partitioned or `separated', relative to the observers.  The term is also, advantageously, {\it a priori} neutral with respect to the concept of locality, and is satisfied by the measurement statistics of all separable quantum states (although there are interesting subtleties as to the precise distinction between equation (1) and so-called `quantum' separability \cite{werner}). 

Summing equation (1) over $a$ and $b$ respectively, for arbitrary $\rho(\lambda)$, it is seen that `separability' immediately implies the weaker property
\begin{equation} \label{nosig}
p(a|x,y,\lambda) =  p(a|x,\lambda) = p(a|x,y',\lambda),~~p(b|x,y,\lambda) =  p(b|y,\lambda) = p(b|x',y,\lambda)  ,
\end{equation}
for the marginal probabilities.  Thus, all locally measurable statistics for each observer are independent of what measurement, if any, might be made by the the other observer.  This property is, these days, typically referred to as `no-signalling'.  For statistical models (where all observable quantities are determined by measurement statistics), it is clearly equivalent to the property that the measurable properties of one system cannot be influenced by operations carried out on the other system. 

Thus, it is equation (2) that, far more appropriately in comparison to equation (1), can be labelled as `locality'.  It is well known that quantum mechanics is local in this sense, where $\lambda$ labels the quantum state (proofs that quantum mechanics satisfies no-signalling, for the general case of physical operations described by linear maps on density operators, appear to have first been given in \cite{hallloc} and \cite{ghirardi}).  Relativistic locality corresponds to the case where $x$ and $y$ are measured in spacelike separated regions.

Finally, it should be noted that the popular term `{\it quantum nonlocality}' carries a totally misleading connotation: that there is something happening faster-than-light in quantum systems.  Yet, as follows from section 2 above, and discussed in greater detail elsewhere \cite{mermin,zukowski,griffiths}, there is in fact no physical justification for such a connotation.  While this term might sound good (and look good on grant applications), this is only because it inherently carries an unjustified suggestion of action at a distance (particularly to non-experts). Mermin concludes its use is no more than `fashion at a distance' \cite{mermin}, and Zukowski notes its appearance in the PACS classification 03.65Ud is not appropriate \cite{zukowski}.  I submit that the label `{\it quantum nonseparability}' would be far less loaded!

\section{Conclusions}

Gisin is able to conclude in \cite{gisin} that quantum mechanics is `nonlocal', irrespective of any consideration of realism, only because his definition of `locality' is formally and empirically equivalent to `local realism', as far as Bell-type inequalities are concerned.  Hence, there are no consequences of his conclusion for debates on locality vs realism - other than as to the question of how `locality' might be physically defined {\it without} implicitly assuming realism.

To clarify such debates, I would like to suggest that the following practices be considered, for the reasons given in sections 2 and 3 above:
\begin{description}
\item[(i)] `locality' for statistical theories, when used in an unqualified sense, be equated with `no-signalling' as per equation (2);
\item[(ii)] the term `local realism' be replaced by `local determinism';
\item[(iii)] the property codified by equation (1) be labelled as `separability'; and 
\item[(iv)] the misleading term `quantum nonlocality' be replaced by `quantum nonseparability'.
\end{description}

Note that (i) and (iii) imply that the Hilbert space model of standard quantum mechanics is both local and
nonseparable.  This model is also nonrealistic - the statistics are nondeterministic (and generated via measures over q-numbers, not c-numbers) - and hence provides an example of a {\it local nonrealistic} model of quantum correlations (in contrast, eg, to `nonlocal realistic' models \cite{bohm}).

Considerations of simplicity, as per the principle of Occam's razor, now lead to a logically rather different possibility than `nonlocality':  given that no-signalling is satisfied by quantum mechanics, why suppose there is some more fundamental `signalling' hidden variable theory? - which would, for example, force one to worry about whether (and why) one can safely ignore what is happening in the Andromeda Galaxy when describing
systems on Earth!  If nature is nonseparable and no-signalling, and quantum mechanics describes these aspects  accurately, why go beyond it?

The above argument emphasises that whatever property one might wish to assume
in addition to no-signalling, to obtain equation (1) - and no matter whether this additional property is called `realism', or `determinism' or `explicability' or `outcome independence' or `classical logic' or `whatever' - it is a property that logically may be given up to explain violations of Bell inequalities.  In particular, there is {\it no} need to assume that anything `faster-than-light' occurs in nature - although this, of course, remains an alternative possibility.

The distinction between `no-signalling' and `separability' is therefore an important topic of investigation in understanding quantum mechanics (assuming this is possible!). It would be fascinating if some corresponding testable inequality could be found.  Note that a further danger of the term `quantum nonlocality' in this regard is that it tends to obscure this distinction - much as Bohr tried to obscure the measurement problem by ignoring the problematic distinction between classical and quantum systems. 

Finally, some (non-original) speculations on the nature of Bell-inequality violations might be in order.  Two possibilities are, of course, a fundamental indeterminism (or `nonrealism'), and faster-than-light signalling. Another is that the current model of spacetime as a simple causal manifold, which provides the empirical justification for no-signalling, is inappropriate (however, it is is not known, for example, how to describe fields without such a model).  Yet another, motivated by Einstein, Podolsky and Rosen \cite{epr}, is `incompleteness' (substantially sharpened as the `outcome independence' property of Jarrett \cite{jarrett}).  Finally, Bohr's reply to Einstein et al.~\cite{bohr}, refusing to permit a physical reality to be ascribed even to a property that could be predicted with probability unity (the property had to actually be measured, via an experimental setup inherently incompatible with other properties), suggests instead the possibility of a fundamental epistemological limit.  

{\bf Acknowledgments:} I thank $\breve{\rm{C}}$aslav Brukner, Arthur Fine, Nicolas Gisin, Robert Griffiths, James Malley and Marek Zukowski for helpful comments.


\newpage
\begin{thebibliography}{99}

 \bibitem{gisin} N. Gisin, eprint arXiv:0901.4255v2 [quant-ph]
 \bibitem{bell} J.S. Bell, Physics {\bf 1} (1964) 195
 \bibitem{bell2} J.S. Bell, Rev. Mod Phys. {\bf 38} (1966) 447 (section V)
 \bibitem{hall} M.J.W. Hall, Int. J. Theoret. Phys. {\bf 27} (1988)
 \bibitem{fine} A. Fine, Phys. Rev. Lett. {\bf 48} (1982) 291 (Propositions 1 and 3)
 \bibitem{fine2} A. Fine, J. Math. Phys. {\bf 23} (1982) 1306 (following Eq.~(11)).  Briefly, if $a^{(m)}_j$ and $b^{(n)}_k$ denote results for measurement pair $(x_m,y_n)$, define hidden variables  $\lambda'_{j_1k_1j_2k_2\dots}:=(a^{(1)}_{j_1},b^{(1)}_{k_1},a^{(2)}_{j_2},b^{(2)}_{k_2},\dots)$; an associated distribution $\rho'(\lambda'):=\int d\lambda \rho(\lambda) p(a^{(1)}_{j_1}|x_1,\lambda) p(b^{(1)}_{k_1}|y_1,\lambda) p(a^{(2)}_{j_2}|x_2,\lambda) p(b^{(2)}_{k_2}|y_2,\lambda)\dots$; and deterministic probabilities $p(a^{(m)}_j|x_m,\lambda'):=1$ ($:=0$) when $j=j_m$ ($j\neq j_m$) for the corresponding $a^{(m)}_{j_m}$ component of $\lambda'$, and similarly for $p(b^{(n)}_k|y_n,\lambda')$.  Since  $\lambda'$ and $\rho'(\lambda')$ depend on the entirety of the particular set of measurement pairs under consideration, the model is nonlocally contextual.
 \bibitem{epr} A. Einstein, B. Podolsky and N. Rosen, Phys. Rev. {\bf 47} (1935) 777
 \bibitem{ch} J.F. Clauser and M. A. Horne, Phys. Rev. D {\bf 10} (1974) 526
 \bibitem{bell3} J.S. Bell, {\it Bertlmann's socks and the nature of reality}, CERN eprint TH.2926, 1980 (reprinted in `Speakable and Unspeakable in Quantum Mechanics', Cambridge University Press, US, 1987)
 \bibitem{jarrett} J.P. Jarrett, No\^{u}s {\bf 18} (1984) 569
 \bibitem{despagnat} B. d'Espagnat, {\it Conceptual Foundations of Quantum Mechanics, 2nd edn} (Benjamin, Reading MA, USA, 1976),  Chapters 8 and 9
 \bibitem{werner} See, eg, R.F. Werner, Phys. Rev. A {\bf 40} (1989) 4277; Zukowksi et al., Phys. Rev. A {\bf 58} (1998) 1694 (and references therein).
 \bibitem{hallloc} M.J.W. Hall, Phys. Lett. A {\bf 125} (1987) 89
 \bibitem{ghirardi} G.C. Ghirardi et al., Europhys. Lett. {\bf 6} (1988) 95
\bibitem{mermin} N.D. Mermin, Found. Phys. {\bf 29} (1999) 571 (\& eprint arXiv:quant-ph/9807055v1)
 \bibitem{zukowski} M. Zukowski, Stud. Hist. Phil. Mod. Phys. {\bf 36} (2005) 566 (\& eprint arXiv:quant-ph/0605034v1) 
 \bibitem{griffiths} R.B. Griffiths, eprint arXiv:0908.2914v1 [quant-ph]
 \bibitem{bohm} D. Bohm, Phys. Rev. {\bf 85} (1952) 166; A.J. Leggett, Found. Phys. {\bf 33} (2003)1469
 \bibitem{bohr} N. Bohr, Phys. Rev. {\bf 48} (1935) 696
 
 
\end{thebibliography}
\end{document}